\documentclass[preprint]{epl}
\usepackage{amsmath}

\makeatletter \@addtoreset{equation}{section} \makeatother
\renewcommand{\theequation}{\arabic{equation}}
\newcounter{alfa}
\newcounter{alfax}
\def\text#1{\mbox{#1}}
\def\equ#1{(\ref{#1})}

\def\bea{\setcounter{alfa}{0}\setcounter{alfax}{1}\addtocounter{alfax}{\value{equation}}
\renewcommand{\theequation}{\addtocounter{alfa}{1}\arabic{alfax}\alph{alfa}}
\begin{eqnarray}}
\def\eea{\end{eqnarray}\setcounter{equation}{\value{alfax}}
\setcounter{alfax}{0}\renewcommand{\theequation}{\arabic{equation}}}
\def\be#1{\begin{equation}\label{#1}}
\def\ee{\end{equation}}
\def\equ#1{(\ref{#1})}

\def\cal{\mathcal}

\def\can#1{{\cal{C}(#1)}}

\title{Equivalence Classes for Gauge Theories}
\author{M\' arcio A. M. Gomes\thanks{Electronic address: gomes@fisica.ufc.br}
\inst{1} and R. R. Landim\thanks{E-mail: \email{renan@fisica.ufc.br}}\inst{1}}
\institute{
  \inst{1} Departamento de
F\'{\i}sica, Universidade Federal do Cear\'a, Caixa Postal 6030,
60455-900, Fortaleza, Cear\'a, Brazil}


\begin{document}

\maketitle

\begin{abstract}
In this paper we go deep into the connection between duality and
fields redefinition for general bilinear models involving the
$1$-form gauge field $A$. A duality operator is fixed based on
"gauge embedding" procedure. Dual models are shown
to fit in equivalence classes of models with the same fields
redefinitions.
\end{abstract}



Beside the importance of duality in understanding various
non-perturbative aspects of field theory and its fundamental role
played in strings theories, several interesting generalizations of
self-dual Chern-Simons-Proca~model \cite{townsend_pk-plb136} and
its equivalent model, \cite{deser_s-plb139}
 the three dimensional topologically massive electrodynamics \cite{schonfeld_jf-npb185,deser_s-prl48,deser_s-ap140},
has been studied in literature. Soon after the work of Deser and Jackiw,
it was realized that self-duality can also occur in Maxwell-Chern-Simons-Proca model
\cite{paul_sk-plb171}. This model has also
been used in the study of bosonization in higher dimensions
\cite{fradkin_e-plb358,banerjee_r-plb358}. Recently it was shown that
there exists a unified theory \cite{ghosh_s-ap291} from which self-dual
model\cite{townsend_pk-plb136}, topologically  massive electrodynamics
\cite{schonfeld_jf-npb185,deser_s-prl48,deser_s-ap140}
and Maxwell-Chern-Simons-Proca systems \cite{paul_sk-plb171} can be
recovered as special cases.

More recently, Lemes {\it et
al}\cite{lemes_ver-plb418,lemes_ver-prd58},  showed that both
abelian and non-abelian Maxwell-Chern-Simons models can be reset
into a pure Chern-Simons term through a suitable local field
redefinition. This was used to evaluate a fermionic determinant
\cite{barci_dg-npb524} and to study the large-mass behavior of the linking-number  in
Maxwell- Chern-Simons theory (MCS) \cite{lemes_ver-jpa32}.  In a
recent paper \cite{gomes_mam-jpa38}, we put the question if two
dual theories do share the same fields redefinition. We showed
that MCS and Maxwell-Proca and their dual models do. A more
profound investigation about this apparently obvious property
reveals that in fact one can construct equivalence classes
embodying all theories with same fields redefinitions, dual models
included.

We begin with the more general bilinear action with a global
symmetry. By construction, it is divided into three parts: a
masslike term, and two terms involving derivatives, one of which
is gauge-invariant and another term that is not. These two sectors
are also shown to be orthogonal to each other. Naturally, the
non-gauge-invariant part drops out in the process of establishing
duality through a "gauge embedding" algorithm \cite{ilha_a-npb604}. We
then present a more direct way to dualize this kind of model by
introducing a duality operator.

Next, we extract group properties from redefining fields by
introducing an abelian group whose action on the connection gauge
field produces another connection gauge field. We call the space
generated by the group action on $A $ "the space of redefined
fields". Gauge theories having the same field redefinition are
collected into classes of equivalence and a criterion is
established to built these classes. Through all paper we lay hold of
functional calculus with differential forms rules introduced in \cite{gomes_mam-jpa38}.

In order to have a better understanding about duality and field
redefinition, we will derive general properties of local bilinear
actions constructed with a gauge field $A$.
\newtheorem{lemma}{Lemma}
\begin{lemma}
The most general local bilinear action constructed with a real
one-form field $A$ is
\be{bilinear}
S=\frac{b_0}{2}(A,A)+\frac{1}{2}(A,\hat{B}A)+\frac{1}{2}(A,\hat{C}A),
\ee where $b_0$ is a constant with canonical dimension
$\can{b_0}=3-2\can{A}$, \bea
&&\hat{B}=\sum_{i=1}^Mb_i(d\ast d\ast)^i,\label{hatB}\\
&&\hat{C}=\sum_{i=1}^Nc_i(\ast d)^i, \eea with
$\can{b_i}=3-2\can{A}-2i$ and $\can{c_i}=3-2\can{A}-i$.
\vskip0.5cm \noindent Proof: \vskip0.5cm A bilinear action on field
$A$ has the general form $(A,\cal{O}A)$, where $\cal{O}$ is an operator
that maps a one-form into another, {\it i.e.} $\cal{O}:\omega_1\rightarrow
\Omega_1$. The operator $\cal{O}$ must be constructed with the
exterior derivative $d$ and the Hodge operator $\ast$. Since
$d^2=0$ and $\ast\ast=\pm1$, there are only two operators in three
dimension that maps a one-form into another: \bea
\ast d:\omega_1\rightarrow \Omega_1,\\
d\ast d\ast:\omega_1\rightarrow \Omega_1,
\eea
hence  $\cal{O}$ has the most general form
\be{formadeO}
\cal{O}=b_0+\sum_{i=1}^Mb_i(d\ast d\ast)^i+\sum_{i=1}^Nc_i(\ast d)^i.
\ee
\end{lemma}
Note that the Laplacian operator $\Delta=dd^\dagger+d^\dagger d$
is a particular case of \equ{formadeO}. The operators $\hat{B}$
and $\hat{C}$ given by \equ{hatB} satisfy the following properties
\bea
&&\hat{B}\hat{C}=\hat{C}\hat{B}=0,\label{BC}\\
&&(\hat{B}\omega_1,\Omega_1)=(\omega_1,\hat{B}\Omega_1),\\
&&(\hat{C}\omega_1,\Omega_1)=(\omega_1,\hat{C}\Omega_1), \eea for
any one-forms $\omega_1$ and $\Omega_1$. The first property of
\equ{BC} is due to the fact that $(\ast d)(d\ast d\ast)=0$ and
$(d\ast d\ast)(\ast d)=0$. The others follow from $(\ast
d\omega_1,\Omega_1)=(\omega_1,\ast d\Omega_1)$ and $(d\ast
d\ast\omega_1,\Omega_1)=(\omega_1,d\ast d\ast\Omega_1)$. $\hat{B}$ and $\hat{C}$
form two orthogonal spaces in sense that
$(\hat{B}\omega_1,\hat{C}\Omega_1)=(\omega_1,\hat{B}\hat{C}\Omega_1)=0$.
Let us observe that if $\omega_1=\hat{C}A$, then
$\delta\omega_1=\hat{C}d\omega_0=0$. Consequently the space
generated by $\hat{C}$ is gauge invariant. Conversely, the space generated by $\hat{B}$
is not gauge invariant.
The $\hat{B}$ term is gauge fixing.
For the bilinear action given by Eq.\equ{bilinear} with
$b_0\ne0$, we can write the operator $\cal{O}$ in a more suitable
way \be{O-new}
\cal{O}=b_0\left(1+\frac{\hat{B}}{b_0}+\frac{\hat{C}}{b_0}\right)=b_0\cal{O}_1^2\cal{O}_2^2,
\ee where \bea
&&\cal{O}_1=\left(1+\frac{\hat{C}}{b_0}\right)^{1/2}=1+\sum_{j=1}^\infty\alpha_j
\left(-\frac{\hat{C}}{b_0}\right)^j,\label{O1}\\
&&\cal{O}_2=\left(1+\frac{\hat{B}}{b_0}\right)^{1/2}=1+\sum_{j=1}^\infty\alpha_j
\left(-\frac{\hat{B}}{b_0}\right)^j,
\eea
with $\displaystyle{\alpha_j=-\frac{(2j)!}{2^{2j}(2j-1)(j!)^2}}$  being the expansion coefficients of the power series.  The  operators $\cal{O}_1$ and $\cal{O}_2$ satisfy the following properties
\bea
&&\left[\cal{O}_1,\cal{O}_2\right]=0,\label{O1O2}\\
&&\left[\ast d,\cal{O}_1\right]=0,\\
&&\left[d\ast,\cal{O}_2\right]=0,\\
&&\ast d\cal{O}_2=\cal{O}_2\ast d=\ast d,\\
&&\cal{O}_1d\ast=d\ast\cal{O}_1 =d\ast. \eea
Using \equ{O-new}, the action in Eq.\equ{bilinear} now reads
\be{bilinear1}
S=\frac{b_0}{2}(\cal{O}_1\cal{O}_2A,\cal{O}_1\cal{O}_2A). \ee We
may wonder what type of gauge invariant action is dual to this
action. The gauge embedding procedure is used. We define the
first iterative action as \be{s1}
S^{(1)}=\frac{b_0}{2}(\cal{O}_1\cal{O}_2A,\cal{O}_1\cal{O}_2A)-(\frac{\delta
S}{\delta A},B), \ee where $B$ is an auxiliary one-form field with
$\delta B=\delta A=d\omega_0$. Then \be{deltaS1} \delta
S^{(1)}=-b_0(\cal{O}_1^2\cal{O}_2^2d\omega_0,B)=-\frac{b_0}{2}\delta(\cal{O}_2B,\cal{O}_2B),
\ee where we used \equ{O1O2}. The next iterative action is gauge
invariant: \be{S2}
S^{(2)}=\frac{b_0}{2}(\cal{O}_1\cal{O}_2A,\cal{O}_1\cal{O}_2A)-b_0(\cal{O}_1\cal{O}_2
A,\cal{O}_1\cal{O}_2B)+\frac{b_0}{2}(\cal{O}_2B,\cal{O}_2B). \ee
Since $\cal{O}_1$ and $\cal{O}_2$ are invertible operators, we can
eliminate the auxiliary field $B$ to rewrite \equ{S2} as a gauge
invariant action depending only on field $A$, \be{seff}
S_{dual}=-\frac{b_0}{2}(\cal{O}_1A,\frac{\hat{C}}{b_0}\cal{O}_1A)=
-\frac{1}{2}(A,\hat{C}A)-\frac{1}{2b_0}(\hat{C}A,\hat{C}A). \ee
This is the dual action of \equ{bilinear1}. Note that the dual
action does not depend on the operator $\hat{B}$. An important
conclusion was found: {\em The dual action depends only on the
gauge invariant sector and the mass term}. In other words a gauge
fixing term is invisible under duality.

We can express the duality map in a more suitable way. From
\equ{bilinear1}, \be{deltaS} \frac{\delta S}{\delta
A}=b_0\cal{O}_1^2\cal{O}_2^2 A, \ee and \equ{seff} can be written
as \be{sdual} S_{dual}=-\frac{1}{2b_0}(\hat{C}A,\frac{\delta
S}{\delta A})=-\frac{1}{2b_0}(\hat{C}A,\frac{\delta }{\delta A})S,
\ee where $\hat{C}\cal{O}_2=\hat{C}$ was used . Then the duality
map is now an operator \be{hatD}
\hat{D}=-\frac{1}{2b_0}(\hat{C}A,\frac{\delta }{\delta A}), \ee
that acts on $S$ giving its dual. For the self-dual model given by
 \begin{equation}
S_{SD}=\int \left( \frac{m^{2}}{2}A{\ast }A+\frac{1}{2}mAdA\right) =\frac{%
m^{2}}{2}\left( A,A\right) -\frac{m}{2}\left( A,{\ast }dA\right),\label{SD}
\end{equation}
the duality operator is \be{hatDsd}
\hat{D}=\frac{1}{2m}(\ast d A,\frac{\delta }{\delta A}). \ee
When $b_0=0$, the best way to find the dual model is writing the operator $\cal{O}$ in terms of the Laplacian and a gauge invariant operator. Since $\Delta=(d\ast)^2-(\ast d)^2$, we have $(d\ast)^2=\Delta+(\ast d)^2$, then 
\be{Odelta}
\cal{O}=\sum_{i=1}^N b_i\Delta^i+\hat{C}'=\hat{B}'+\hat{C}',
\ee
where $\hat{C}'$ is the gauge invariant part. Then applying the same procedure we find the dual action,
\be{dual-b0}
S_{dual}=-\frac{1}{2}(A,\hat{C}'A)-\frac{1}{2}(\hat{C}'A,\hat{C}'\hat{B}'^{-1}A)=\hat{D}S,
\ee
where 
\be{nonlocaldual}
\hat{D}=-\frac{1}{2}(\hat{C}'A,\hat{B}'^{-1}\frac{\delta }{\delta A}),
\ee
is a non-local  duality operator, since $\hat{B}'^{-1}$ is non-local. In this work we are only taking into account the local operators.

Let us go back to local operator $\cal{O}_1$. Its action  on a gauge field $A$ produces another
gauge field with the same gauge transformation, i.e, if $\delta
A=d\omega_0$, then $\delta \cal{O}_1A=d\omega_0$. Let us denote this
set of operators by $G_1$.
   An element $g_1$ of $G_1$ has the form
\[g_1=(1+\hat{P}),\]
with $\hat{P}d=0$ and $d\ast\hat{P}=0$. We shall prove that $G_1$
is an abelian group. In fact, for $g_1$ and $g'_1$ $\in$ $G_1$,
\[g_1g'_1=(1+\hat{P})(1+\hat{P}')=(1+\hat{P}+\hat{P}'+\hat{P}\hat{P}')=(1+\hat{P}'')=g''_1
\in G_1,\] since
\[(\hat{P}+\hat{P}'+\hat{P}\hat{P}')d=\hat{P}d+\hat{P}'d+\hat{P}\hat{P}'d=0\]
and
\[d\ast(\hat{P}+\hat{P}'+\hat{P}\hat{P}')=d\ast\hat{P}+d\ast\hat{P}'+d\ast\hat{P}\hat{P}'=0.\]
The inverse and identity elements are well defined:
\[(1+\hat{P})^{-1}=(1+\sum_{n=0}^{\infty}(-1)^n\hat{P}^n)=(1+\hat{P}_{-1}),\quad
(1+\hat{P})(1+\hat{P}_{-1})=1.\]
 The action of $G_1$ on $A$ defines a space of redefined fields. Any redefined field has
 the same gauge transformation of $A$. An element of the space of redefined fields
 is $\hat{A}=(1+\hat{P})A$. A redefined field is completely specified by the action
 of an element of $G_1$. If $\hat{A}=(1+\hat{P})A$ and $\hat{A}'=(1+\hat{P}')A$,
 then $\hat{A}=\hat{A}'\Longleftrightarrow\hat{P}=\hat{P}'$.

 For a gauge invariant theory the bilinear action is
 \be{inv4}
 S=\frac{1}{2}(\hat{C}A,A),
 \ee
 with $\hat{C}$ being a polynomial in $\ast d$ with at least one zero root.
 We can always express $\hat{C}$ in the form
 \be{hatC4}
 \hat{C}=c_0(\ast d)^jg^2_1,
 \ee
 for some $j \in N$ and $g_1\in G_1$ with $c_0$ a constant. Then any gauge invariant
 bilinear action can be written as
 \be{rede4}
 S=\frac{c_0}{2}(\hat{A},(\ast d)^j\hat{A}),
 \ee
where $\hat{A}=g_1A=(1+\hat{P})A$ is a redefined field. Let us
observe that $(\ast d)^j g_1$ can not belongs to $G_1$. Therefore
another gauge invariant action can present the same redefined
field . For example
\[
S'=\frac{c'_0}{2}(\hat{A},(\ast d)^{k}\hat{A}),
\]
for $k\ne j$, has the same redefined field  of \equ{rede4}, but it
is a different gauge theory. $S$ and $S'$ are related since they
have the same redefined field. We are now ready to construct a
relation between these two gauge theories.
\newtheorem{definition}{Definition}
\begin{definition}
Two gauge actions $S$ and $S'$ are said to belong to the same equivalence class
if there exist $p$ and $p'$ $\in N$  and a constant $a_0\ne0$ such that
\be{relation}
 a_0(\ast d)^p\frac{\delta S}{\delta A}=(\ast d)^{p'}\frac{\delta S'}{\delta A}.
 \ee
We shall denote this relation by $\sim$.
\end{definition}
  It is easy to see that this relation $\sim$ defined above is indeed an equivalence relation:
  i.e. (i) $S\sim S$ (reflexive); (ii) if $S\sim S'$, then $S'\sim S$ (symmetric); and (iii)
  if $S\sim S'$ and $S'\sim S''$, then $S\sim S''$ (transitive).
  Let us verify the last one. If  $S\sim S'$ and $S'\sim S''$, then there are numbers
  $p,p'$ and $q',q''$ and constants $a_0$ and $ b_0$ such that
\[ a_0(\ast d)^p\frac{\delta S}{\delta A}=(\ast d)^{p'}\frac{\delta S'}{\delta A}, \]
\noindent and
\[b_0(\ast d)^{q'}\frac{\delta S'}{\delta A}=(\ast d)^{q''}\frac{\delta S''}{\delta A}.\]
Then
\[a_0b_0(\ast d)^{p+q'}\frac{\delta S}{\delta A}=(\ast d)^{p'}b_0(\ast d)^{q'}
\frac{\delta S'}{\delta A}=(\ast d)^{p'+q''}\frac{\delta
S''}{\delta A},\] implies that $S\sim S''$. The actions that
fulfill condition \equ{relation} form an equivalence class of
gauge theories. We now prove that $S$ and $S'$ belonging to same
equivalence class must have the same field redefinition. To see this,
let $S$ and $S'$ given by
\[S=\frac{c_0}{2}(\hat{A},(\ast d)^j\hat{A}),\]
\noindent and
\[S'=\frac{c'_0}{2}(\hat{A'},(\ast d)^k\hat{A'}),\]
be two actions that have in principle different field
redefinitions. Since by hypothesis they belong to the same
equivalent class, $\exists$~ $p$, $p'$ and a constant $a_0$ such
that
\[a_0c_0(\ast d)^{j+p}g^2_1A=c'_0(\ast d)^{k+p'}g'^2_1A,\]
holds. Now using the fact that $g^2_1=1+\hat{P}$ and $g'^2_1=1+\hat{P}'$ we have
\[a_0c_0(\ast d)^{j+p}A+a_0c_0(\ast d)^{j+p}\hat{P}A=c'_0
(\ast d)^{k+p'}A+c'_0(\ast d)^{k+p'}\hat{P}'A,\] which implies
that $a_0c_0=c'_0$, $j+p=k+p'$ and $\hat{P}=\hat{P}'$.
Consequently, $\hat{A}=\hat{A}'$. Any member of a class has the
same redefined field and the same solution of the equation of motion.
For instance, the self-dual and Maxwell-Chern-Simons actions are
members of the same class, since they have the same redefined
field. This class may be called the class of self-dual model.
These two models considered are also dual to each other. This is
not coincidence. Looking at equation \equ{sdual}, we can see that
$S_{dual}$ and $S$ belong to the same class if $\hat{C}=c_0(\ast
d)^j$. This is just the case of the self-dual model with
$\hat{C}=-m\ast d$.

Another interesting equivalence class is obtained from the
Maxwell-Proca action \be{proca}
S_{MP}=\frac{m^2}{2}(A,A)-\frac{1}{2}(\ast dA,\ast
dA)=\frac{m^2}{2}(A,\left(1-(\ast d/m)^2\right)A) \ee In this case
$\hat{C}=-(\ast d)^2$, and the group element is
\[g_1=(1-(\ast d/m)^2)^{1/2}=1+\sum_{j=1}^\infty\alpha_j(\ast d/m)^{2j}.\]
The Maxwell-Proca action and its dual action belong to the same
class \be{procadual} S_{dual-MP}=\frac{1}{2m^2}\left((\ast
d)^2A,\frac{\delta S_{MP}}{\delta A}\right)=\frac{1}{2}(\ast
dA,\ast dA)-\frac{1}{m^2}(\ast d\ast dA,\ast d\ast dA), \ee or in
terms of redefined fields \be{proca-redef}
S_{MP}=\frac{m^2}{2}(\hat{A},\hat{A}), \ee and
\be{procadual-redef} S_{dual-MP}=\frac{1}{2}(\ast d\hat{A},\ast
d\hat{A}). \ee

Let us underline that if two models belongs to the same class, its
bilinear terms  have in common an invertible operator given by
\equ{O1}. This means that the propagators are related in some way.
We know that if two models are dual they represent same physics in
different scales of energy. We can argue that if two models belong
to the same equivalence class, they represent the same physics in
certain scales of energy. This point will be analyzed in a future
work.

An important point is how to construct the non-abelian version of our results. For this, we would begin from the requirement  that the field redefinition must be local and the redefined field remains a connection under BRST transformation, namely
\be{conection}
s\hat{A}=dc+g[\hat{A},c]=\hat{D}c,
\ee
where $\hat{A}=A+\sum_{i=1}^{\infty}A_i/m^i$, $m$ is a mass parameter and $g$ is the coupling constant.
It follows that
\be{brst}
sA_i=g[A_i,c].
\ee
Since that the cohomology of the non-abelian version is isomorphic to its abelian counterpart, the knowledge of the abelian redefinition allows us to discover the non-abelian expansion. A detailed analysis will be given in a future work.

In conclusion, in this paper we study the connection between
duality and field redefinition. We show that a general bilinear
three-dimensional action for the $1$-form gauge field $A$ can be
rewritten as a unique term with a suitable field redefinition. We
construct equivalence classes for models sharing same field
redefinition. Dual models belong to the same class. Another
important issue is the derivation of a duality operator.

\vskip0.3cm
  \noindent
  {\large\bf Acknowledgments}
  \vskip0.2cm
Conselho Nacional de Desenvolvimento Cient\'\i fico e
tecnol\'ogico-CNPq is gratefully acknowledged for financial
support.

\end{document}